\documentclass[12pt]{article}
\usepackage{putex}
\usepackage{graphicx}

\begin{document}

\preprint{PUPT-2255}

\institution{PU}{Joseph Henry Laboratories, Princeton University, Princeton, NJ 08544}

\title{Breaking an Abelian gauge symmetry near a black hole horizon}

\authors{Steven S. Gubser}

\abstract{I argue that coupling the Abelian Higgs model to gravity plus a negative cosmological constant leads to black holes which spontaneously break the gauge invariance via a charged scalar condensate slightly outside their horizon.  This suggests that black holes can superconduct.}

\PACS{}
\date{January 2008}

\maketitle

\section{Introduction}

In \cite{Gubser:2005ih} it was suggested that black hole horizons could exhibit spontaneous breaking of an Abelian gauge symmetry if gravity were coupled to an appropriate matter lagrangian, including a charged scalar that condenses near the horizon.  Depending on parameters, it was further suggested that a flux lattice might arise, providing a particularly stark example of the limitations of no-hair theorems.  (For a review of no-hair theorems, see for example \cite{Bekenstein:1996pn}.)  The matter lagrangian suggested in \cite{Gubser:2005ih} was somewhat complicated, involving two Abelian gauge fields and a non-renormalizable coupling of the scalar to one of them.  Here it will be argued that spontaneous symmetry breaking arises near black hole horizons for theories as simple as this one:
 \eqn{GravityPlusAbelianHiggs}{
  16\pi G_N {\cal L} = R - {6 \over L^2} - {1 \over 4} F_{\mu\nu}^2 - 
    |\partial_\mu \psi - i q A_\mu \psi|^2 - m^2 |\psi|^2 \,,
 }
where I use mostly plus signature and define the matter fields in such a way that Newton's constant $G_N$ enters as an overall prefactor.  The first two terms in \eno{GravityPlusAbelianHiggs} are the Einstein-Hilbert lagrangian plus a negative cosmological constant, which means that the simplest solution to the equations of motion is anti-de Sitter space.  I will focus on four spacetime dimensions, although the mechanism to be described seems likely to work in other dimensions.  Recent work including \cite{Herzog:2007ij,Hartnoll:2007ih,Hartnoll:2007ip,Hartnoll:2008hs} has suggested the possibility that charged black holes in four-dimensional anti-de Sitter space ($AdS_4$) may provide useful analogies to phenomena observed in layered or thin-film superconductors above the transition temperature, in particular the large Nernst effect observed in \cite{OngNature} and later works.

The remaining terms in \eno{GravityPlusAbelianHiggs} are the Abelian Higgs lagrangian, except that the potential is missing the usual $|\psi|^4$ term.  This term can be added without changing the story much.  A novelty relative to the usual Ginzburg-Landau story is that one can keep $m^2$ constant and positive and still get the symmetry breaking to happen.  The way it happens is that an electrically charged black hole makes an extra contribution to the scalar potential which makes the $\psi=0$ solution unstable, provided that $q$ is large enough, and that $m^2$ is not too positive, and that the black hole is sufficiently highly charged and sufficiently cold.  To see this, let's consider configurations of the form
 \eqn[c]{GeneralBH}{
  ds^2 = g_{tt} dt^2 + g_{rr} dr^2 + ds_2^2  \cr
  A_\mu dx^\mu = \Phi dt \qquad \psi = \psi(r) \,,
 }
where all fields are assumed to depend only on $r$.  The ``transverse'' metric $ds_2^2$ could be proportional to the metric on a unit two-sphere, or it could be proportional to the flat metric $d\vec{x}^2$ on ${\bf R}^2$.  Suppose $\psi$ is too small to back-react significantly upon the geometry.  To investigate the effective potential experienced by the scalar, one may restrict attention to the last two terms in \eno{GravityPlusAbelianHiggs}.  They are
 \eqn{JustPsi}{
  16\pi G_N {\cal L}_\psi = -g^{tt} q^2 \Phi^2 |\psi|^2 - 
    g^{rr} |\partial_r \psi|^2 -
    m^2 |\psi|^2 \,.
 }
So the effective mass of $\psi$ is
 \eqn{meff}{
  m_{\rm eff}^2 = m^2 + g^{tt} q^2 \Phi^2 \,.
 }
Because $g^{tt}$ is negative outside the horizon and diverges to $-\infty$ near the horizon, it seems inevitable that $m_{\rm eff}^2$ should become negative there if $\Phi$ is non-zero.  In fact, the story is slightly more subtle: as I explain at the end of section~\ref{PERTURB}, for $\psi$ to be non-zero at the horizon, one must choose a gauge where $\Phi=0$ there.\footnote{It is argued in \cite{Kobayashi:2006sb} that $\Phi$ must vanish at a horizon even if one doesn't consider non-zero $\psi$: roughly, the argument is that the one-form $\Phi dt$ is ill-defined at the horizon because $dt$ has infinite norm there.}  If $q$ is big, and the electric field outside the horizon is big (meaning that $\Phi$ rises very quickly from $0$ to finite values), and $m^2$ is small, then it seems clear that $m_{\rm eff}^2$ can still become negative a little outside the horizon.  It is less clear that it is possible for $m_{\rm eff}^2$ to be negative enough for long enough to produce an unstable mode in $\psi$.  What I want to argue is that this is possible in asymptotically $AdS_4$ geometries.  The actual computation I will do is to find marginally stable perturbations around solutions that do not break the $U(1)$ symmetry.  Such marginally stable modes exist on co-dimension one surfaces of parameter space provided $q$ is big enough.  Logically speaking, this is not quite enough to guarantee the existence of an instability for sufficiently small temperatures.  But it is highly suggestive.

It has long been understood that extremal black holes can exhibit perfect diamagnetism \cite{Wald:1974np,Bicak:1980du,Chamblin:1998qm}, but the reasons for it are essentially geometrical, whereas the proposals of \cite{Gubser:2005ih} and of the current paper focus on a charged scalar condensate that develops slightly outside the horizon of non-extremal black holes.  Interaction of black holes with charged scalar condensates and flux vortices have also been considered previously \cite{Chamblin:1997gk}, but in contexts where the gauge symmetry is broken at asymptotic infinity, whereas in \cite{Gubser:2005ih} and in the present paper, it is only broken near the horizon.

The organization of the rest of this paper is as follows.  In section~\ref{RNADS} I present the charged black hole backgrounds of interest and discuss two scaling symmetries associated with them.  Section~\ref{PERTURB} summarizes the marginally stable linearized perturbations around these solutions that break the $U(1)$ symmetry.  Conclusions and directions for future work are summarized in section~\ref{CONCLUDE}.

\section{The Reissner-Nordstrom black hole in $AdS_4$}
\label{RNADS}

The electrically charged black hole in $AdS_4$ is
 \eqn[c]{RNAdS}{
  ds^2 = -f dt^2 + {dr^2 \over f} + r^2 d\Omega_{2,k}^2
   \qquad\hbox{where}\qquad f = k - {2M \over r} + 
     {Q^2 \over 4r^2} + {r^2 \over L^2}  \cr
  \Phi = {Q \over r} - {Q \over r_H} \qquad\qquad \psi = 0 \,.
 }
It is a solution to the equations of motion following from the lagrangian \eno{GravityPlusAbelianHiggs}.  In \eno{RNAdS}, $d\Omega_{2,k}^2$ is a metric of constant curvature, with scalar curvature equal to $2k$.  If $k=0$, then $d\Omega_{2,k}^2$ is the line element of flat ${\bf R}^2$.  If $k>0$, then $d\Omega_{2,k}^2$ is the metric of a two-sphere, $S^2$, of radius $1/\sqrt{k}$.  If $k<0$ (a case I will not consider below), then $d\Omega_{2,k}^2$ is the metric of the hyperbolic plane with radius of curvature $1/\sqrt{-k}$.\footnote{It is an interesting case, though: quotienting the hyperbolic plane by a discrete subgroup, one can obtain smooth surfaces of any genus $g>1$.  A superconducting medium on such a surface has some interesting topological excitations.}  The asymptotically $AdS_4$ region is at large $r$.  A regular black hole horizon occurs if $f$ has a positive root, let's say at $r=r_H$.  Let's assume that a horizon exists and regard $M$ as determined by $r_H$ and the other parameters in \eno{RNAdS} through the equation $f(r_H)=0$.  The parameters $M$ and $Q$ are not precisely the mass and the charge of the black hole; rather, they are quantities with dimensions of length to which the mass and charge (or mass density and charge density if $k=0$) are related through factors of $G_N$ and/or $k$.

Before entering into a discussion of perturbations around \eno{RNAdS}, it is helpful to consider two scaling symmetries.  Let's say that a quantity $X$ has a charge $\alpha$ under a particular scaling symmetry if its scale transformation is
 \eqn{Xcharge}{
  X \to \lambda^\alpha X \,.
 }
Then the two scaling symmetries of interest are characterized by the following charge assignments:
 \begin{equation}
 \begin{tabular}{c||c|c|c|c|c||c|c|c||c|c|c|c}
  & $t$ & $r$ & $k$ & $M$ & $Q$ & $ds^2$ & $\Phi dt$ & $\psi$ & $L$ & $m$ & $q$ & $G_N$ \\ \hline
  first symmetry & $1$ & $-1$ & $-2$ & $-3$ & $-2$ & $0$ & $0$ & $0$ & $0$ & $0$ & $0$ & $0$ \\
  second symmetry & $1$ & $1$ & $0$ & $1$ & $1$ & $2$ & $1$ & $0$ & $1$ & $-1$ & $-1$ & $2$
 \end{tabular}\label{ScalingSymmetries}
 \end{equation}
We're always interested in having a black hole horizon, so $r_H \neq 0$; also, we're always interested in having non-zero charge, so $Q \neq 0$.  So an efficient use of the scaling symmetries \eno{ScalingSymmetries} is to set $r_H=Q=1$.  Then one sees that the solutions \eno{RNAdS} modulo the scaling symmetries are a two-parameter family, labeled by $k$ and $L$.  One can pass to the flat space limit by taking $L \to \infty$ with $k$ finite, and one can pass to the Poincar\'e patch of $AdS_4$ by taking $k \to 0$ with $L$ finite.

\section{Marginally stable modes}
\label{PERTURB}

A marginally stable perturbation is one where $\psi$ depends only on $r$ and is infinitesimally small (that is, it doesn't back-react on the other fields).  The main interest attaches to the first marginally stable mode that appears as one lowers the temperature, which corresponds to raising $L$ if one fixes $r_H=Q=1$ as described following \eno{ScalingSymmetries}.  The simplest possibility is for the appearance of this mode to signal a second order phase transition to an ordered state where $\psi$ is non-zero.  I will assume that $\psi$ is everywhere real, because phase oscillations in the $r$ direction would only raise the energy of the mode, making it less likely to become unstable.

The equation of motion for $\psi$ following from the lagrangian \eno{JustPsi} is
 \eqn{psiEOM}{
  {1 \over \sqrt{-g}} \partial_r \sqrt{-g} g^{rr} \partial_r \psi + 
    m_{\rm eff}^2 \psi = 0 \,,
 }
where $m_{\rm eff}^2$ is as specified in \eno{meff}.  The factors of $\sqrt{-g}$ arise because the action is $S = \int d^4 x \, \sqrt{-g} {\cal L}$.  It is straightforward to show that \eno{psiEOM} takes the following form:
 \eqn{psiLinear}{
  \psi'' &+ {-1 + (8r-4)k + 4 (4r^3-1)/L^2 \over 
    (r-1) (-1 + 4 kr + 4 r (r^2+r+1)/L^2)} \psi'  \cr
    &\qquad{} + 
    {m_{\rm eff}^2 \over (r-1) (-1 + 4 kr + 4 r (r^2+r+1)/L^2)}
     \psi = 0 \,,
 }
where I have set $r_H=Q=1$, and
 \eqn{meffAgain}{
  m_{\rm eff}^2 = m^2 - {4q^2 (r-1) \over
   -1 + 4 k r + 4 r (r^2+r+1)/L^2} \,.
 }
Evidently, \eno{psiLinear} now depends on four parameters: $k$, $L$, $m$, and $q$.  They are subject only to the constraint
 \eqn{LkConstraint}{
  -1 + 4k + 12/L^2 \geq 0 \,,
 }
which guarantees that $r=1$ is the largest root of $f$ and that the denominators in \eno{psiLinear} and~\eno{meffAgain} are positive for $r>1$.\footnote{If one tries to go to values of $L$ and $k$ that violate the inequality \eno{LkConstraint}, there may still be a horizon, just for some value $r>1$.  When this happens, it doesn't mean that the solutions are new solutions.  Instead, it means that the scaling symmetry should be used anew to send the horizon to $r=1$.  Then one will recover solutions in a range of parameters specified in \eno{LkConstraint}.}  When the inequality \eno{LkConstraint} is saturated, the black hole is extremal in the sense of having zero temperature.

\begin{figure}
  \centerline{\includegraphics[width=3in]{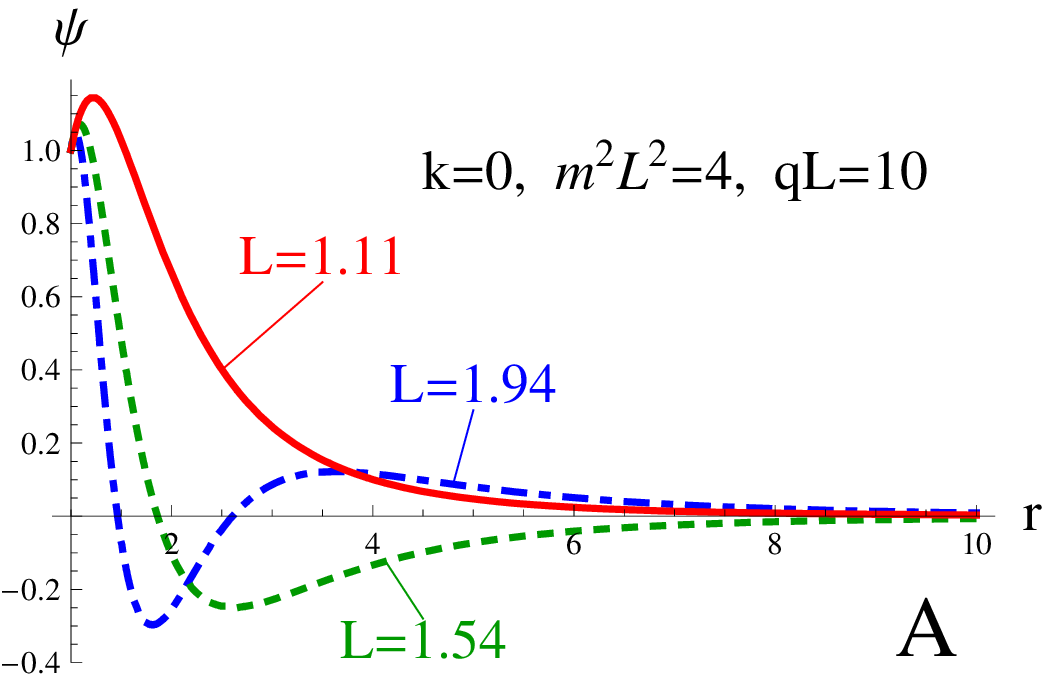}}\ \\
  \centerline{\includegraphics[width=3in]{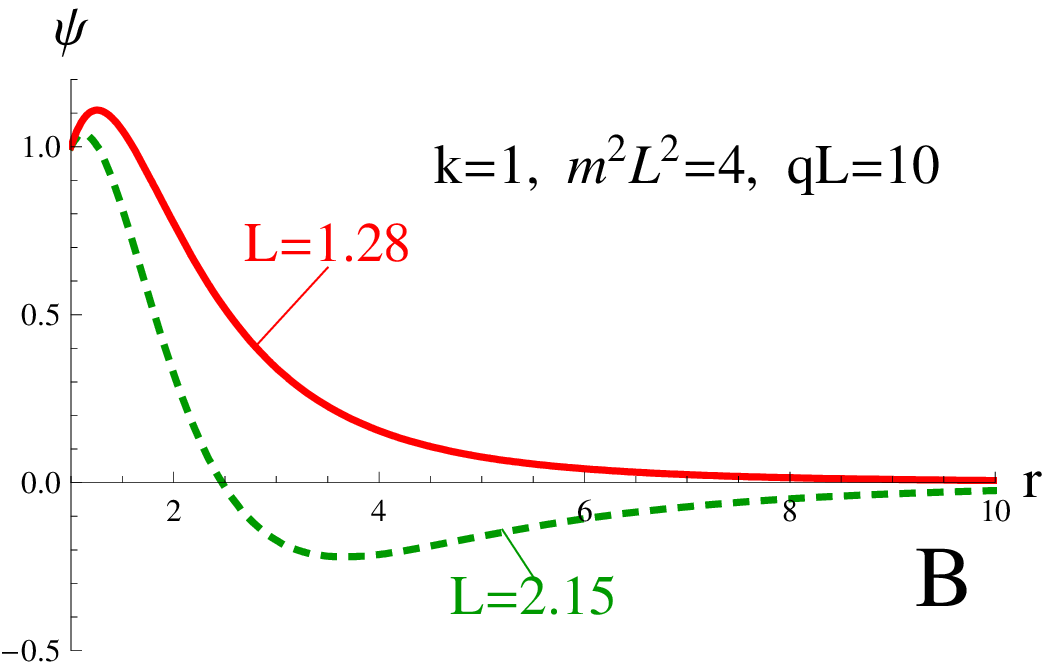}}\ \\[10pt]
  \centerline{\includegraphics[width=3in]{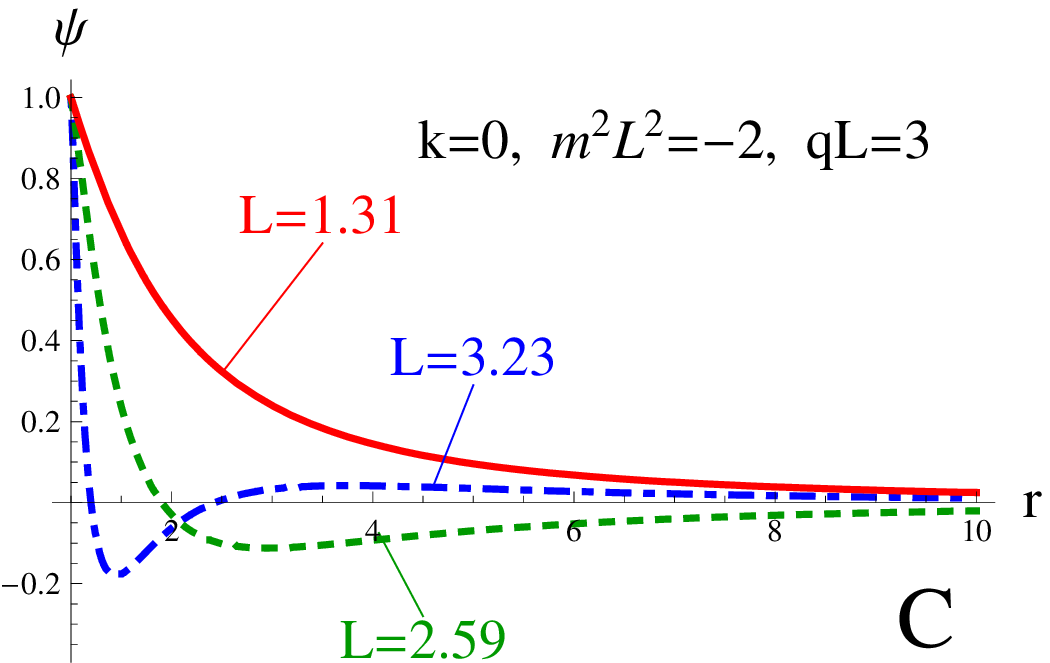}}
  \caption{Examples of marginally stable modes in an asymptotically $AdS_4$ geometry.}\label{AdSModes}
 \end{figure}

Finding a marginally stable mode comes down to solving a boundary value problem.  One boundary is the horizon, and the other is asymptotic infinity (i.e.~either anti-de Sitter space or flat space).  The horizon is at $r=1$, which is a regular singular point of the differential equation \eno{psiLinear}.  The leading order solutions there are
 \eqn{HorizonSolutions}{
  \psi = \psi_0 + \tilde\psi_0 \log (r-1) \,,
 }
where $\psi_0$ and $\tilde\psi_0$ are arbitrary integration constants.  The appropriate horizon boundary condition is $\tilde\psi_0=0$, because otherwise $\psi$ is singular.  Let's take advantage of the linearity of \eno{psiLinear} to set $\psi_0=1$, keeping in mind that it would really be more proper to treat $\psi_0$ as a parameter whose smallness justifies the neglect of back-reaction.  One can easily develop $\psi$ in a series solution: the first three terms are
 \eqn{SeriesSoln}{
  \psi(r) = 1 + {4m^2 \over -1 + 4k + 12/L^2} (r-1) - 
    {2 ((1+12/L^2) m^2 - 2m^4 + 2q^2) \over
     (-1 + 4k + 12/L^2)^2} (r-1)^2 + O[(r-1)^2] \,.
 }
The boundary condition at asymptotic infinity is simple if $m^2>0$: one must have $\psi \to 0$.  In $AdS_4$, $m^2 < 0$ is also an interesting case, because then $\psi$ corresponds to some relevant operator in the dual conformal field theory.  But specifying the appropriate behavior near the conformal boundary in this case involves some technical issues that I prefer to postpone.  If $m^2>0$, then the presence of marginally stable modes obviously can only be due to the second term in \eno{meff}.  The following choice of parameters provides a good example of the phenomenon:
 \eqn{NiceParams}{
  k=0 \qquad m^2 L^2 = 4 \qquad qL = 10 \,.
 }
I prefer to hold $m^2 L^2$ and $qL$ fixed as I vary $L$ because these combinations are invariant under the scaling symmetries \eno{ScalingSymmetries}.  What this means is that I hold the theory fixed (up to scalings which affect none of the classical equations of motion) while varying $L$.  Alternatively, one could hold $L$ and $Q$ fixed while varying $r_H$: that is, keep charge density fixed as one lowers the entropy and temperature.  As shown in figure~\ref{AdSModes}A, there are multiple marginally stable modes at different values of $L$.  For the parameter choice indicated in \eno{NiceParams}, there appears to be an infinite tower of these modes at larger and larger values of $L$ (meaning lower and lower temperature if one prefers to hold $L$ and $Q$ fixed instead of $r_H$ and $Q$), converging to the limiting value $L = 2\sqrt{3}$ at which the black hole becomes extremal.  The marginally stable mode that occurs for the smallest value of $L$, call it $L=L_*$, is the important one, as it can be expected to be unstable for $L>L_*$.  Probably this mode condenses, and the other modes do not matter to the phase structure of the black hole.  Non-zero $k$ doesn't change the story much, as shown in figure~\ref{AdSModes}B, except that, for the choice of parameters studied, there appear to be only two marginally stable modes.

The story in $AdS_4$ does not appear to change qualitatively when $m^2 < 0$, provided one satisfies the Breitenlohner-Freedman bound \cite{Breitenlohner:1982bm}, $m^2 L^2 > -9/4$, which guarantees that perturbations of $\psi$ don't make the $AdS_4$ vacuum unstable.  But there is one complication, namely the choice of boundary condition at the conformal boundary.  The asymptotic behavior of solutions of \eno{psiLinear} at large $r$ is
 \eqn{LargeR}{
  \psi \sim {A_\psi \over r^{3-\Delta}} + {B_\psi \over r^\Delta} \,,
 }
where $\Delta$ is determined by
 \eqn{DeltaDef}{
  m^2 L^2 = \Delta (\Delta-3) \,.
 }
According to the gauge-string duality \cite{Maldacena:1997re,Gubser:1998bc,Witten:1998qj}, when $\psi$ is associated with an operator ${\cal O}_\psi$ of dimension $\Delta$, one should choose $A_\psi=0$ to describe states of the theory in the absence of a deformation of the lagrangians by ${\cal O}_\psi$.  The difficulty is that there are two solutions to \eno{DeltaDef}, namely
 \eqn{DeltaPM}{
  \Delta_\pm = {3 \pm \sqrt{9+4m^2L^2} \over 2} \,,
 }
and switching from $\Delta_+$ to $\Delta_-$ effectively swaps $A_\psi$ and $B_\psi$.  When $m^2 L^2 > -5/4$, one finds $\Delta_- < 1/2$, which is disallowed by unitary.  But for $-9/4 < m^2 L^2 < -5/4$, either solution to \eno{DeltaDef} is permitted by unitarity, and choosing one or the other corresponds to selecting one of two possible boundary theories, distinguished by the dimension of ${\cal O}_\psi$ \cite{Klebanov:1999tb}.  I considered the case $m^2 L^2 = -2$, which is in this window, and I chose the large $r$ asymptotics corresponding to an expectation value of an operator with dimension $\Delta=2$ (i.e.~I used $\Delta_+$ not $\Delta_-$).  See figure~\ref{AdSModes}C.  Unsurprisingly, it turns out that a smaller value of $q$ suffices to tip the system into an instability well away from extremality---which is to say, well above zero temperature.

The story changes drastically in the limit $L \to \infty$ with $k=1$, which describes a Reissner-Nordstrom black hole in asymptotically flat spacetime.  Consider the effective mass in this case:
 \eqn{meffFlat}{
  m_{\rm eff}^2 = m^2 - 4q^2 + {4 (-1+4k) q^2 r \over -1 + 4kr} \,,
 }
where as usual I have used the scaling symmetries to set $r_H=Q=1$.  The bound \eno{LkConstraint} now translates to $k \geq 1/4$, and for this range of $k$, the last term in \eno{meffFlat} is seen to be monotonically decreasing with $r$.  So if $\psi$ is stable at infinity, it is stable at the horizon.  Therefore, no near-horizon symmetry breaking is possible in this case.

It is interesting to consider the extremal limit $k=1/4$ in asymptotically flat space a bit more carefully.  One sees from \eno{meffFlat} that $m_{\rm eff}^2$ is constant, but it could be positive or negative depending on whether $m$ is bigger or smaller than $2q$.  What this indicates is that if a particle is so highly charged that its gravitational attraction to the black hole is overcome by its electrostatic repulsion, then there is an instability: the black hole super-radiates the highly charged particles to bring its charge-to-mass ratio down until the charged particles are no longer repelled from it.  One might ask why this doesn't happen in the asymptotically $AdS_4$ examples: after all, the value $qL=10$ that I used when $m^2 L^2 = 4$ is pretty big.  The difference is that massive particles cannot escape from $AdS_4$, no matter how highly charged they are: the conformal boundary is inaccessible to particles with $m^2>0$.  A heuristic view of what happens is that the highly charged quanta try to escape, fail, and so have no choice but to condense inside $AdS_4$.  This intuition doesn't work so well when $m^2<0$, because tachyons can reach the boundary of $AdS_4$.  The boundary conditions at the conformal boundary then become important, and one can loosely think of the one I chose in the discussion below \eno{DeltaDef} as perfectly reflecting.

\begin{figure}
  \centerline{\includegraphics[width=3in]{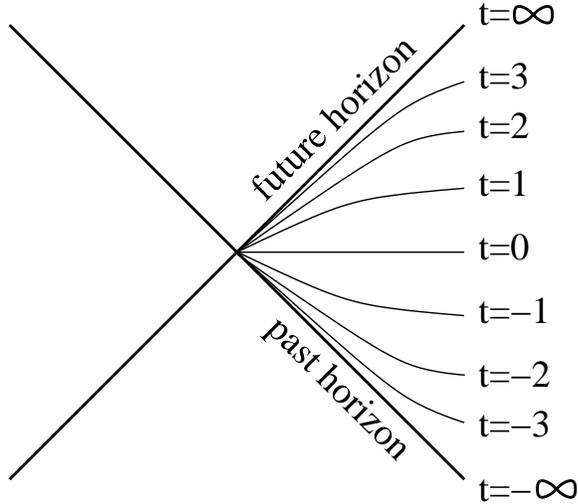}}
  \caption{The Penrose diagram of a static black hole horizon.  Each point in the regions on the right and the top represents a two-dimensional slice at constant $t$ and $r$ of the metric \eno{RNAdS}.  The other two regions arise from continuing the metric to a spacetime whose only failures to be geodesically complete arise from curvature singularities.  The region on the right connects to asymptotic infinity and corresponds to $r>r_H$.  Contours of constant $t$ are shown in this region.}\label{Penrose}
 \end{figure}

Lastly, let's turn to the issue of why a preferred gauge choice is to set $\Phi=0$ at the horizon when $\psi$ is non-zero there.  Suppose I made a different choice: $\Phi=\Phi_H$ at the horizon.  Because of the term $-g^{tt} |\partial_t \psi - i q \Phi \psi|^2$ in the lagrangian, the energy of a scalar configuration with $\psi \neq 0$ at the horizon will be infinite unless $\psi$ has a time-dependence $\psi \propto e^{iq\Phi_H t}$.  It does not seem so bad to repeat the analysis of this section in such a gauge.  But consider the global spacetime.  As shown in figure~\ref{Penrose}, the future horizon is at $t=+\infty$, the past horizon is at $t=-\infty$, and the intersection of the two is where $t$ runs from $-\infty$ to $+\infty$.  This illustrates in a standard way the fact that $t$ is not a good coordinate at the horizon.  The trouble with $\psi \propto e^{iq\Phi_H t}$ is that $\psi$ is then ill-defined on the horizon---that is, on the past and future horizons as well as their intersection.

\section{Conclusions}
\label{CONCLUDE}

Modulo two logical gaps (points~\ref{Marginal} and~\ref{Entropic} in the list below) I have shown that the Abelian Higgs model exhibits spontaneous symmetry breaking near the horizon of electrically charged black holes in $AdS_4$, provided the charges of the black hole and the scalar are large enough and their masses are small enough.  This suggests that black hole horizons can superconduct when they get cold enough---although it would be more precise to say that a superconducting layer forms slightly outside the horizon.  There are two known mechanisms by which such spontaneous symmetry breaking can occur.  One was suggested in \cite{Gubser:2005ih}, and it hinges on a non-renormalizable coupling in the matter lagrangian.  This mechanism can be made to work in asymptotically flat space, and the charge of the scalar field can be small.  The mechanism described in this paper employs only renormalizable interactions among the matter fields.  (Gravity, of course, is non-renormalizable in the usual sense.)  But it doesn't work in flat space, so the four-part no-hair conjecture proposed at the end of \cite{Gubser:2005ih} is not jeopardized.

It is interesting to inquire whether there are any lessons from gravity about what it takes to superconduct in a field theory dual to an asymptotically $AdS_4$ background.  If one relies entirely on the mechanism described in this paper (as opposed to the one suggested in \cite{Gubser:2005ih}) then the main feature of interest is that the charge $q$ of the scalar field needs to be large enough.  This charge is not to be confused with the charge of a Cooper pair; nor is the gauge field $F_{\mu\nu}$ to be confused with the one associated with photons.  Instead, one must rely upon the translation of $AdS_4$ to field theory quantities provided by the gauge-string duality.  The gauge field in $AdS_4$ is associated via this dictionary with a conserved current $J_m$ in the field theory, and this current is the electrical current.  (Roman indices like $m$ run over the $2+1$ dimensions of the boundary theory, while Greek indices like $\mu$ run over the $3+1$ dimensions of $AdS_4$.)  The complex scalar $\psi$ is associated with a complex operator ${\cal O}_\psi$ in the field theory, which most naturally would be the operator that destroys a Cooper pair.  What $q$ controls is an overall factor on the three-point function $\langle J {\cal O}_\psi {\cal O}_\psi^* \rangle$.  More accurately, $q$ controls a ratio of three-point and two-point functions:
 \eqn{OOJ}{
  {\langle J {\cal O}_\psi {\cal O}_\psi^* \rangle^2 \over
   \langle J J \rangle 
    \langle {\cal O}_\psi {\cal O}_\psi^* \rangle^2} \propto q^2 \,.
 }
To phrase this even more precisely, one could take advantage of the fact that conformal invariance plus conservation of $J$ dictates the functional form of all the correlators appearing in \eno{OOJ}, up to overall normalization factors and possibly contact terms.\footnote{Of course, what I mean is that conformal invariance controls the form of the zero-temperature limit of the correlators of interest.  At finite temperature, the discussion here can be understood as applying to correlators measured on length scales significantly smaller than the inverse temperature.}  A ratio analogous to \eno{OOJ} can be formed from the overall normalization factors, and it is that ratio which is controlled by $q$.  So the punchline is that this ratio needs to be sufficiently large in order for spontaneous symmetry breaking to take place by the mechanism described in this paper.

The mechanism suggested in \cite{Gubser:2005ih} involved two gauge fields and was formulated with asymptotically flat space in mind, but it seems clear that it can be made to work in $AdS_4$.  Also, it may be possible to set the gauge fields equal and use only one.\footnote{I thank C.~Herzog for discussions on these points.}  The gauge coupling can be small---in fact the calculation as presented in \cite{Gubser:2005ih} is well-controlled only in the limit of weak gauging.  The crucial non-renormalizable interaction is $\ell^2 |\psi|^2 F_{\mu\nu}^2$, and the coefficient $\ell^2$ has to be sufficiently large in order for spontaneous symmetry breaking to occur.  Presumably there is some relation analogous to \eno{OOJ} of roughly the form
 \eqn{OOJJ}{
  {\langle JJ {\cal O}_\psi {\cal O}_\psi^* \rangle \over
    \langle JJ \rangle \langle {\cal O}_\psi {\cal O}_\psi^* 
      \rangle} \propto \ell^2 \,.
 }
It is more complicated to give precise meaning to \eno{OOJJ} than to \eno{OOJ}: conformal invariance does not entirely fix the form of four-point functions, and from the gravity side, both the non-renormalizable coupling and the renormalizable gauge coupling would contribute to the four-point function in question.  Nevertheless, one sees again in \eno{OOJJ} the theme that mixed correlators of $J$ and ${\cal O}_\psi$ have to be large enough compared to the two-point functions of these operators in order for spontaneous symmetry breaking to occur.

Let's end with an assessment of two logical gaps in the arguments of this paper (points \ref{Marginal} and \ref{Entropic} below); a point on which I feel some genuine confusion (point~\ref{RealWorld}); and several directions for future work (points~\ref{Explore}, \ref{Superconduct}, and~\ref{Miscellaneous}):
 \begin{enumerate}
  \item Does the existence of a marginally stable mode imply an instability?  Generically the answer is yes: if one continuously varies parameters, encountering a marginally stable fluctuation usually means that on one side or the other, the fluctuation becomes unstable.  If one relies on this genericity argument, then the unstable mode has to be on the side of low temperatures, because the scaling symmetries relate the high-temperature (large $r_H$) limit with the $Q \to 0$ limit, where it seems obvious that no instability can occur.\label{Marginal}
  \item Are solutions with $\psi \neq 0$ entropically favored over the ones with $\psi=0$?  It seems like common sense that they should be, because the unstable modes (assuming they are unstable) want to condense, and area theorems say they can do so only if the endpoint (assuming there is a static endpoint) has higher entropy.  But there is no substitute for an explicit calculation on this point.  For example, in the microcanonical ensemble, one could compare the entropy of broken and unbroken solutions at fixed energy and charge.\label{Entropic}
  \item My exploration of when marginally stable modes arise was anything but exhaustive.  The differential equation \eno{psiLinear} depends on four parameters, but instead of trying to scan the full parameter space, I looked along just three one-dimensional curves inside it.  Interesting threshold effects might exist for special parameter choices where marginal modes barely exist.  The aim of this paper has been to give a proof of principle (modulo points~\ref{Marginal} and~\ref{Entropic}) that the mechanism described below \eno{meff} really works; further work is needed to establish the full range of behaviors that this mechanism exhibits.\label{Explore}
  \item Does the breaking of the gauge invariance of the lagrangian \eno{GravityPlusAbelianHiggs} really mean that the dual theory superconducts?  Recall that $F_{\mu\nu}$ isn't the gauge field associated with real world photons!  This is not likely to be a problem, because non-zero $\psi$ translates via the gauge-string dictionary into an expectation value for ${\cal O}_\psi$, and such an expectation value would break the $U(1)$ gauge invariance of a gauge field in the boundary theory under which ${\cal O}_\psi$ is charged.  To be really sure that superconductivity occurs, one should compute a two-point correlator $\langle JJ \rangle$ in the broken phase.  This is a harder calculation than the ones presented here because one must first construct a solution in the broken phase of the non-linear equations following from \eno{GravityPlusAbelianHiggs} and then perturb the gauge field (and maybe the other fields too) around it.  Through such a calculation, one could in principle access the frequency and wave-vector dependence of $\langle JJ \rangle$, not just the transport coefficients available from its infrared behavior.\label{Superconduct}
  \item Do black holes have anything to do with real superconductors?  Unlike the situation with quantum chromodynamics, where there is a good understanding from D-brane constructions of how one gets interacting gluons, the duality of gravitational systems and strongly correlated electron systems is not, as yet, supported by a first principles construction in which the underlying degrees of freedom (e.g.~the electrons) are manifest.  So I'm not sure how seriously to take the analogies between black holes and superconducting materials presented to date, including the one in this paper.  Naturally, I hope that black holes might provide some useful hints about the correct description of high~$T_c$ superconductivity in a regime that standard field theory techniques cannot access.  But this is only a hope.\label{RealWorld}
  \item Can the infrared dynamics of the condensate whose existence I have argued for be described by an effective theory like Ginzburg-Landau?  Does it lead to a type I or a type II superconductor?  Is the transition to superconductivity first order or second order?  Can one construct flux vortices explicitly?  Can one calculate, or estimate, $H_{c1}$ and $H_{c2}$?  Do the results of such calculations bear any quantitative resemblance to real-world superconductors?  Can the mechanism described here be realized in a string theory or M-theory construction?\label{Miscellaneous}
 \end{enumerate}
I look forward to addressing some of these issues in future work.

\section*{Acknowledgments}

This work was supported in part by the Department of Energy under Grant No.\ DE-FG02-91ER40671 and by the NSF under award number PHY-0652782.  I thank C.~Herzog for useful discussions and for bringing references \cite{Kobayashi:2006sb,Chamblin:1998qm,Chamblin:1997gk} to my attention.

\bibliographystyle{ssg}
\bibliography{higgs}
\end{document}